\documentclass[conference]{IEEEtran}
\IEEEoverridecommandlockouts

\usepackage{cite}
\usepackage{amsmath,amssymb,amsfonts}
\usepackage{algorithmic}
\usepackage{multirow}
\usepackage{booktabs}
\usepackage{graphicx}
\usepackage[table,xcdraw]{xcolor}
\usepackage{textcomp}
\usepackage{xcolor}
\def\BibTeX{{\rm B\kern-.05em{\sc i\kern-.025em b}\kern-.08em
    T\kern-.1667em\lower.7ex\hbox{E}\kern-.125emX}}

\begin{document}

\title{Improved Feature Extraction Network for Neuro-Oriented Target Speaker Extraction\\

\thanks{
$^*$Corresponding Author

This work is supported by the {STI 2030—Major Projects (No. 2021ZD0201500)}, the National Natural Science Foundation of China (NSFC) (No.62201002, 6247077204), Excellent Youth Foundation of Anhui Scientific Committee (No. 2408085Y034), Distinguished Youth Foundation of Anhui Scientific Committee (No. 2208085J05), Special Fund for Key Program of Science and Technology of Anhui Province (No. 202203a07020008), Cloud Ginger XR-1.}
}

\author{
\IEEEauthorblockN{\textit{Cunhang Fan$^{1}$, Youdian Gao$^{1}$, Zexu Pan$^{2}$, Jingjing Zhang$^{2}$, Hongyu Zhang$^{1}$,Jie Zhang$^{3}$,Zhao Lv$^{1,}$$^*$}}
\IEEEauthorblockA{$^{1}$School of Computer Science and Technology, Anhui University, Hefei, China}{$^{2}$National University of Singapore, Singapore}\\
{$^{3}$NERC-SLIP, University of Science and Technology of China (USTC), Hefei, China}\\
}

\maketitle







\maketitle

\begin{abstract}
 The recent rapid development of auditory attention decoding (AAD) offers the possibility of using electroencephalography (EEG) as auxiliary information for target speaker extraction. However, effectively modeling long sequences of speech and resolving the identity of the target speaker from EEG signals remains a major challenge. In this paper, an improved feature extraction network (IFENet) is proposed for neuro-oriented target speaker extraction, which mainly consists of a speech encoder with dual-path Mamba and an EEG encoder with Kolmogorov-Arnold Networks (KAN). We propose SpeechBiMamba, which makes use of dual-path Mamba in modeling local and global speech sequences to extract speech features. In addition, we propose EEGKAN to effectively extract EEG features that are closely related to the auditory stimuli and locate the target speaker through the subject's attention information. Experiments on the KUL and AVED datasets show that IFENet outperforms the state-of-the-art model, achieving 36\% and 29\% relative improvements in terms of scale-invariant signal-to-distortion ratio (SI-SDR) under an open evaluation condition.

\end{abstract}


\section{Introduction}
In a multi-speaker environment, speech is often affected by noise and interference speech signals, which is known as the cocktail party problem. Speech separation\cite{c46,c47,c48} and target speaker extraction\cite{c39,c40,c36,c37,c38} algorithms are proposed to solve this problem.

Target speaker extraction differs from speech separation techniques in the capacity of filtering without prior knowledge of speaker count and resolving the pervasive global permutation ambiguity. 
The former leverages a reference speech sample from the target speaker, mimicking the human brain's top-down attentional mechanism. By directing this focused attention with the aid of the reference speech, the system efficiently extracts the target speech. Beyond mere reference speech, innovative works have explored the incorporation of spatial location cues\cite{c3}, context-aware comprehension\cite{c4}, and video information\cite{c5,c35} as auxiliary references. 
Nevertheless, a notable limitation persists: these methods fall short of understanding the real human brain’s attention, i.e., the neuronal responses, in the cocktail party.

\begin{figure*}[htbp]
\centering 
\centerline{\includegraphics[width=18.3cm,height=8.3cm ]{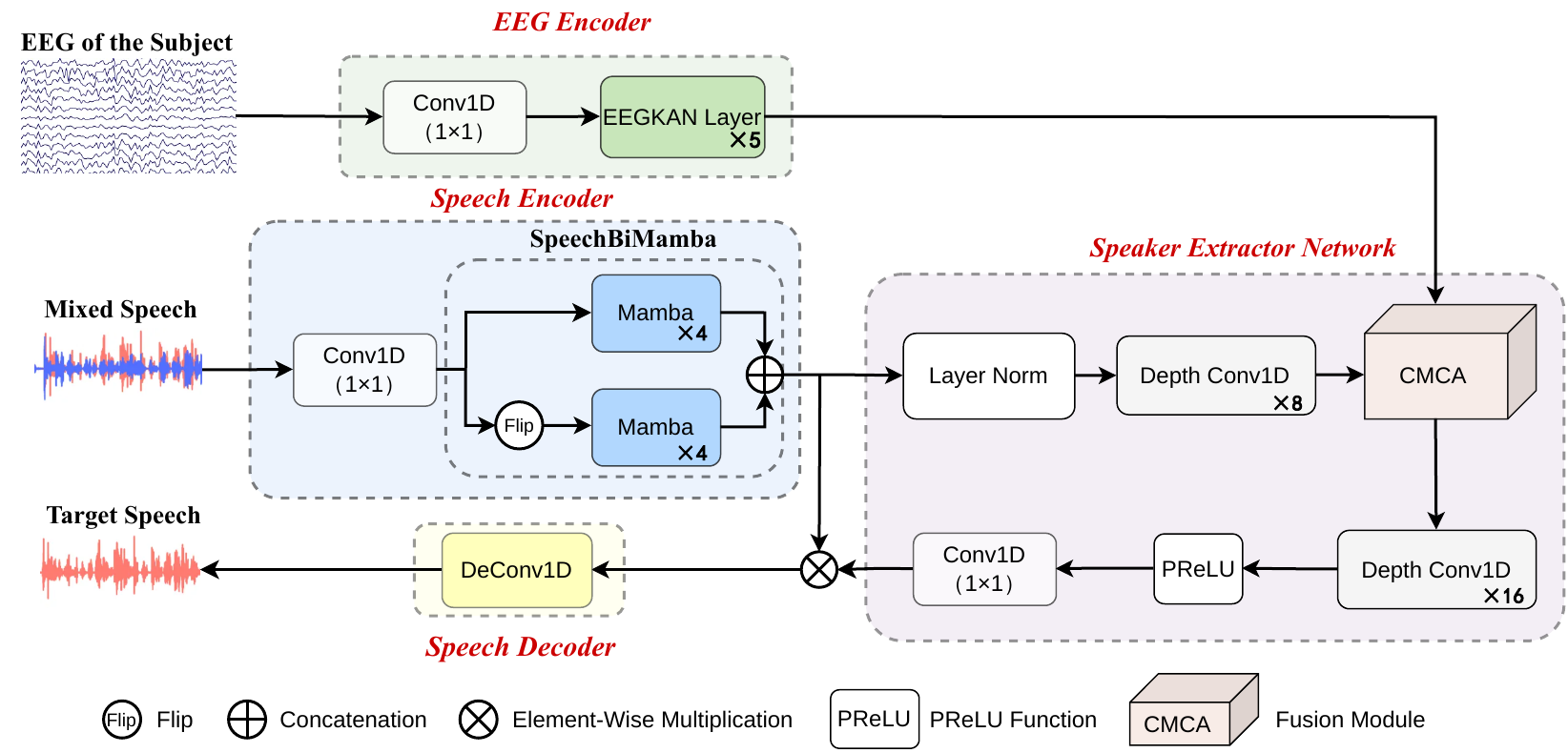}}
\caption{The overall structure of IFENet.}
\label{fig}
\end{figure*}

To further simulate the auditory perception of the human brain, the researchers decode the speech streams the subjects are paying attention to from the electroencephalography (EEG) signals, known as auditory attention decoding (AAD)\cite{c30,c44,c45,c49,c50,c51}. In recent years, research has intensified on neural-oriented (EEG cue) target speaker extraction based on AAD. This advancement has significantly propelled the evolution of hearing aids, cochlear implants, and other applications. The brain-informed speech separation (BISS)\cite{c6} model does not use EEG signals directly as input, but instead uses speech envelopes reconstructed from EEG signals as reference. BESD\cite{c7} and U-BESD\cite{c8} use an end-to-end architecture, execute entirely in the time domain, and fuse EEG signals using feature-wise linear modulation. BASEN\cite{c9} uses the deep convolutional network to extract EEG embeddings and uses cross-attention to fuse EEG signals and speech. The neuro-steered speaker extraction network NeuroHeed\cite{c23} and NeuroHeed+\cite{c27} use the attention mechanism and an AAD backend to better model the EEG information in a time-domain end-to-end architecture. MSFNet\cite{c10} fuses speech features at different time scales and EEG features. However, since speech is a signal that changes continuously with time, it is difficult to capture this long-distance dependence by using only a one-dimensional convolutional network to extract the features of speech. Moreover, EEG contains a lot of interference components, only convolutional networks or deep convolutional networks can not extract deeper speaker-related information in EEG signals, and the receptive field is limited, so the computational complexity will be very high.

 This paper proposes a time-domain improved feature extraction network (IFENet) for neural-oriented target speaker extraction. We introduce SpeechBiMamba to help extract speech features and model long sequences of speech through the structure of dual-path Mamba. SpeechBiMamba makes use of a dual-path network to model speech sequences locally and globally and process them forward and backward. A new EEGKAN structure is also proposed in the EEG encoder to extract the information of the EEG signal better and to determine the target speaker according to the implied attention information in the EEG by improving multiple attention (MHA)\cite{c11,c28}. The extracted speech embeddings and EEG embeddings are then fused in a separate network and the corresponding masks are estimated. Finally, the mask is restored to the final speech waveform. We conduct experiments on two datasets, the publicly available KULeuven (KUL) dataset\cite{c31}, and the laboratory-acquired AVED dataset\footnote{http://iiphci.ahu.edu.cn/toAuditoryAttention}. On the KUL and AVED datasets, The proposed IFENet achieves a relative improvement of 36\% and 29\% over the MSFNet in SI-SDR\cite{c12}, and the perceptual evaluation of speech quality (PESQ)\cite{c13} increases by 6.6\% and 11.7\%.

\section{Proposed Model}
\label{sec:pagestyle}

\subsection{Overall Architecture}
\label{ssec:subhead}

Our proposed model IFENet is shown in Fig. 1. The clean speech of the target speaker is extracted with the aid of EEG signals. IFENet is based on ConvTasNet\cite{c14} and follows an end-to-end architecture with the main components being a speech encoder, an EEG encoder, a speaker extractor network, and a speech decoder. By fusing speech with EEG to extract comprehensive features, the target speech is extracted through the subject's attention. Each part is described as follows.

Given a mixed speech signal, the speech encoder converts the speech signal into speech embedding, the EEG encoder encodes the EEG signal of the N channels. Speech embedding and EEG embedding are fused in the speaker extractor network. Notably, we use the convolutional multi-layer cross attention (CMCA) module proposed in \cite{c9} for feature fusion, which consists of multiple layers of cross-attention blocks with jump connections and normalization between each two layers. The speech and EEG features of the left and right branches are added layer by layer, and then spliced with the original speech embeddings and EEG embeddings to obtain the fused features. 
Then the corresponding mask is estimated (only the target speech is allowed to pass), and the speech decoder maps the mask to get the reconstructed speech signal.

\subsection{Speech Encoder}
\label{ssec:subhead}
\subsubsection{Mamba}
\label{sssec:subsubhead}

Mamba\cite{c15} is based on a state-space model (SSM)\cite{c16} which can perform well in the modeling of long sequences. The specific description is as follows:
\begin{equation} 
{h}' \left ( t \right ) =Ah\left ( t \right ) +Bx\left ( t \right ),
\end{equation} 
\begin{equation} 
y(t)=Ch(t)
\end{equation}
SSM performs sequence-to-sequence mapping, where the state matrix $A\in R^{N\times N}$, $B, C\in R^{N}$ are parameters and $h\in R^{N}$ is the hidden state. Motivated by SSM, Mamba introduces two unique mechanisms, as shown in Fig. 2(a), Mamba incorporates a unique selection mechanism (dependent on input) and a hardware-aware algorithm that is linear to sequence length to help quickly cycle the model through a single scan. Owing to its outstanding performance and straightforward architecture, has demonstrated remarkable superiority across diverse tasks\cite{c41,c42,c43}. Specifically, in the field of speech applications, Mamba has exhibited comparably advanced capabilities, as evidenced by studies \cite{c32,c33,c34}. Furthermore, it boasts significant computational efficiency, exemplified by innovations such as dual-path Mamba \cite{c20} and TF-Mamba \cite{c21}.

\subsubsection{SpeechBiMamba}
\label{sssec:subsubhead}
As shown in Fig. 1, SpeechBiMamba makes use of dual-path Mamba to model local and global speech sequences and extract speech features. Specifically, we first pass the input mixed speech through the Conv1D, downsampling the speech to extract the basic features of the speech. After that, the original shape and the flipped shape of the extracted speech features are processed through the dual-path Mamba structure. The original speech and the flipped speech undergo multiple transformations via the Mamba block, which is executed N times (N=4 in this study), and then the output is connected in series to get the speech features after Mamba modeling.

\begin{figure}[htbp]
\centering 
\centerline{\includegraphics[width=9cm, height=6.82cm]{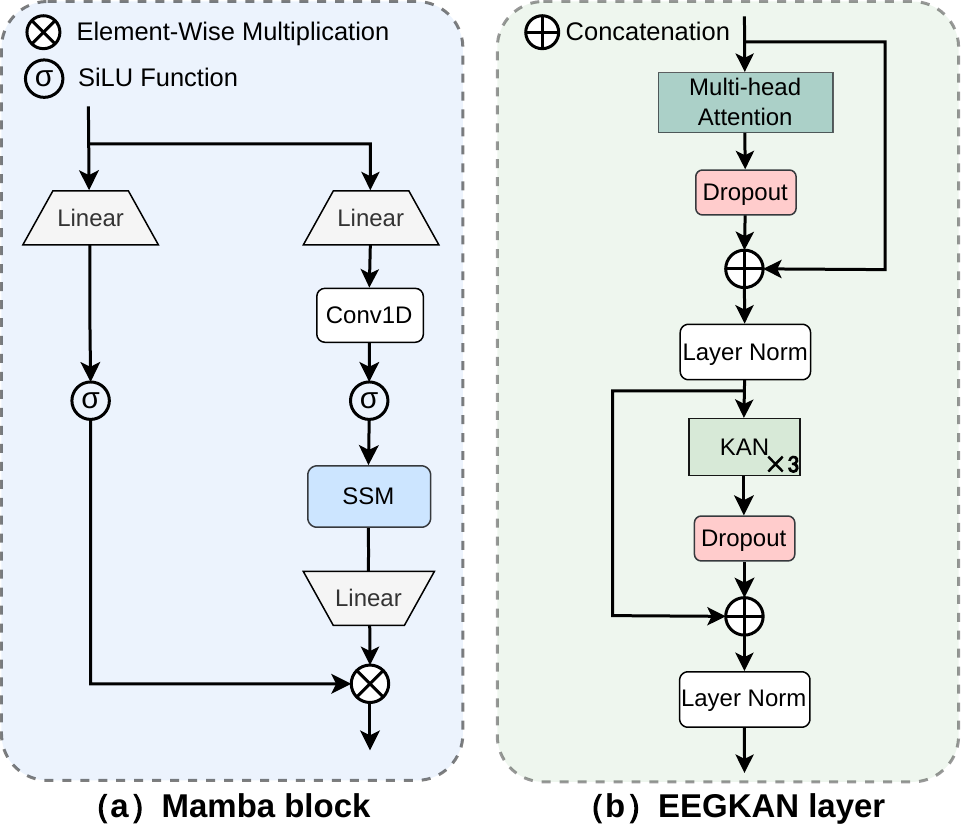}}
\caption{(a) Mamba block, (b) EEGKAN layer.}
\label{fig}
\end{figure}

\subsection{EEG Encoder}
\label{ssec:subhead}
\subsubsection{KAN}
\label{sssec:subsubhead}
 The proposal of Kolmogorov-Arnold Networks (KAN)\cite{c22} is inspired by the Kolmogorov-Arnold representation theorem, and KAN has the same fully connected structure as MLPs. However, unlike traditional MLPs, MLPs set fixed activation functions on the node, and KAN places learnable activation functions on the side. At the same time, KAN nodes do not need to perform any nonlinear transformation of the input signal, only need to perform simple summation. More importantly, KAN is superior to MLPs in terms of accuracy and interoperability.

\subsubsection{EEGKAN}
\label{sssec:subsubhead}

Inspired by Neuroheed\cite{c23} and Kansformer\cite{c24}, we notice that features of EEG signals can be better extracted using attentional mechanisms, and Kansformer can enhance the nonlinear feature representation and interpretability of the model. Therefore, we propose a novel EEG encoder based on the multi-head attention mechanism and using KAN instead of MLPs. The specific structure is shown in Fig. 2(b). The module in turn consists of a multi-head attention block with two attention heads, Dropout (0.5), LayerNorm, three KAN layers, Dropout (0.5), and LayerNorm.

\subsection{Loss Function}
\label{ssec:subhead}

In our work, we use the negative SI-SDR as the loss function, which is often used for speaker extraction with good performance. The SI-SDR is defined as:
\begin{equation}  
SI\text{-}SDR=10\log_{10}\frac{{\left \| \frac{ \hat{s} ^{T}s}  {{\left \| s  \right \| }^{2}} s  \right \| } ^{2} }{  \left \| \frac{  \hat{s} ^{T}s}  {{\left \| s  \right \| }^{2}} s  -\hat{s} \right \|  ^{2} }
\end{equation} 
where \(\hat{s}\) and \( \mathit{s} \) represent extracted speech signal and clean speech signal respectively. The performance of the model was evaluated by maximizing the SI-SDR between $\hat{s}$ and $s$ as the training goal. The higher the SI-SDR is, the better the quality of the speech signal is.

\section{Experimental setup}
\label{sec:typestyle}

\subsection{Datesets}
\label{ssec:subhead}

\subsubsection{KUL dataset}
\label{sssec:subsubhead}
 The dataset consists of 16 subjects with normal hearing, each of which performs 20 trials. We use the first 8 trials in which each subject participated, in which subjects are presented with different speech in the left and right ears. Participants are asked to pay attention to the sounds in one ear and ignore the sounds in the other.  The BioSemi ActiveTwo system is used to record 64-channel EEG signals at an 8196 Hz sample rate. 

\subsubsection{AVED dataset}
\label{sssec:subsubhead}
We propose a new dataset for tasks related to auditory attention decoding which includes 20 subjects with normal hearing, with an average age of 20, all of whom sign informed consent forms. Each subject conduct 16 consecutive trials. Similar to the KUL dataset, each subject wear in-ear headphones and is asked to pay attention to speech on one side. All of the speech comes from Mandarin stories told by a man and a woman. 32-channel EEG data are recorded at a rate of 1kHz.

\subsection{Training and Testing Setups}
\label{ssec:subhead}

 For both datasets, we downsample the EEG data to 128Hz and set the speech sampling rate to 44.1kHz. Due to the uniqueness of EEG, the EEG of each subject is very different. To make full use of the EEG information, we train and test a model on the data of each subject, and the final result is the average of all subjects. So we divide each trial of a subject into a training set, a verification set, and a test set, with the proportions of 80\%, 10\%, and 10\%, respectively, with no overlap of speech stimuli between the groups.

\subsection{Training Details}
\label{ssec:subhead}

We use PyTorch to conduct our experiments. All of our model training is done on an NVIDIA GeForce RTX 4090 GPU. On the KUL dataset, all models are trained with 300 epochs and a batch size of 8, using the Adam optimizer. The maximum learning rate is 0.0002, the learning rate adjustment strategy is linear preheating under cosine annealing, and the preheating ratio is 5\%. On the AVED dataset, all models are trained with 400 epochs, and the remaining settings are consistent with those on the KUL dataset.



\begin{table}[]
\centering
\caption{The comparison of various methods on the KUL dataset and AVED dataset. BASEN$^{*}$ is the BASEN model that uses a multi-head attention module for the EEG encoder.}

\resizebox{0.5\textwidth}{!}{  
\addtolength{\tabcolsep}{2.5pt}
\begin{tabular}{@{}ccccc@{}}
\toprule
\multirow{2}{*}{Models}           & SI-SDR        & STOI          & ESTOI         & PESQ          \\ \cmidrule(l){2-5} 
                                  & \multicolumn{4}{c}{KUL}                                       \\ \midrule
BASEN\cite{c9}   & 3.51          & 0.85          & 0.75          & 2.43          \\ 
BASEN$^{*}$      & 6.30          & 0.86          & 0.76          & 2.39          \\ 
MSFNet\cite{c10} & 5.05          & 0.84          & 0.74          & 2.29          \\ \midrule
IFENet (ours)                     & \textbf{6.85} & \textbf{0.87} & \textbf{0.77} & \textbf{2.44} \\ \midrule
                                  & \multicolumn{4}{c}{AVED}                                      \\ \cmidrule(l){2-5} 
BASEN\cite{c9}   & 6.58          & 0.84          & 0.72          & 1.76          \\ 
BASEN$^{*}$     & 6.75          & 0.85          & 0.71          & 1.72          \\ 
MSFNet\cite{c10} & 6.78          & 0.85          & 0.71          & 1.71          \\ \midrule
IFENet (ours)                     & \textbf{8.76} & \textbf{0.88} & \textbf{0.77} & \textbf{1.91} \\ 
\bottomrule
\end{tabular}
}
\end{table}

\subsection{Evaluation Metrics}
\label{ssec:subhead}

 We mainly use three indicators to evaluate the effectiveness of the method, including SI-SDR in dB, PESQ, and short-term objective Intelligibility (STOI)\cite{c25}. Among them, SI-SDR is often used in speech separation tasks and has good robustness. PESQ and STOI are used to evaluate the perceived quality and intelligibility of extracted speech relative to unprocessed multi-speaker speech signals. In addition to this, we also consider extended short-term objective intelligibility (ESTOI) as a supplement.

\section{Results}
\label{sec:majhead}

To verify the validity of IFENet, we compare the performance of different models. In addition, we also conduct ablation experiments to verify the effects of our proposed EEGKAN and SpeechBiMamba blocks.

\subsection{Comparison with Baseline Models}
\label{ssec:subhead}

In TABLE \uppercase\expandafter{\romannumeral1}, we compare IFENet's test results with those of different advanced models, including BASEN, BASEN$^{*}$ and MSFNet, in which BASEN$^{*}$ is the BASEN model that uses multi-head attention module for the EEG encoder. On the KUL dataset, we can see that IFENet has a 36\% improvement compared with MSFNet on SI-SDR and is superior to BASEN, BASEN$^{*}$ and MSFNet on PESQ, STOI, and ESTOI. Besides, BASEN$^{*}$ performs quite well, with SI-SDR only 0.55 dB lower than IFENet and higher than MSFNet. This again proves the role of the attention mechanism in EEG feature extraction. 

In addition, IFENet on the AVED dataset performs the best, which is 1.98 dB higher than MSFNet on SI-SDR, with a 29\% improvement, outperforms MSFNet by 0.2 in PESQ, 0.3 in STOI, and 0.6 in ESTOI. BASEN$^{*}$ also performs well, coming close to MSFNet on all metrics, but below IFENet. Therefore, we can conclude that the IFENet model still exhibits competitive performance compared to other neuro-oriented target speaker extraction under different datasets and experimental Settings.

\subsection{Ablation Study}
\label{ssec:subhead}

To illustrate the effects of our proposed SpeechBiMamba and EEGKAN respectively, we compare the methods with or without SpeechBiMamba and EEGKAN in TABLE \uppercase\expandafter{\romannumeral2}.  

Obviously, we find the EEGKAN to be more effective than the SpeechBiMamba. On the KUL dataset, IFENet without EEGKAN is much worse than IFENet without SpeechBiMamba. On the AVED dataset, similar to the KUL dataset, IFENet without SpeechBiMamba is still stronger than that of IFENet without EEGKAN, SI-SDR and PESQ achieve an increase of 1.23 dB and 0.05 respectively. This finding confirms that both EEGKAN and SpeechBiMamba are beneficial for speaker extraction.

Furthermore, our analysis reveals that the combined utilization of the SpeechBiMamba and the EEGKAN yields the most pronounced enhancement.  On the KUL dataset, IFENet achieves remarkable results, with an SI-SDR of 6.85 dB, STOI of 0.87, and PESQ of 2.44. Although the PESQ  marginally trails the 2.45 achieved when employing solely the SpeechMamba block, it still demonstrates good performance.  Similarly, in the AVED dataset, IFENet stands out as the superior model across various metrics, notably attaining an SI-SDR of 8.76 dB, STOI of 0.88, and PESQ of 1.91. These findings prove the significance of harnessing the profound information in both speech and EEG signals, which we achieve by integrating the EEGKAN and SpeechBiMamba blocks.

\begin{table}[]

\caption{Ablation experiment of EEGKAN and SpeechBiMamba on the KUL dataset and AVED dataset. w/o means without.}
\resizebox{0.5\textwidth}{!}{  
\scalebox{1}{

\addtolength{\tabcolsep}{-1.8pt}

\begin{tabular}{@{}ccccc@{}}
\toprule
\multirow{2}{*}{Models} & SI-SDR        & STOI          & ESTOI         & PESQ          \\ \cmidrule(l){2-5} 
                        & \multicolumn{4}{c}{KUL}                                       \\ \midrule
IFENet                  & \textbf{6.85} & \textbf{0.87} & \textbf{0.77} & 2.44          \\
w/o EEGKAN              & 3.89          & 0.85          & 0.76          & \textbf{2.45} \\
w/o SpeechBiMamba       & 6.64          & 0.86          & 0.77          & 2.38          \\ \midrule
                        & \multicolumn{4}{c}{AVED}                                      \\ \midrule
IFENet                  & \textbf{8.76} & \textbf{0.88} & \textbf{0.77} & \textbf{1.91} \\
w/o EEGKAN              & 7.38          & 0.87          & 0.75          & 1.85          \\
w/o SpeechBiMamba       & 8.61          & 0.87          & 0.77          & 1.90          \\ \bottomrule
\end{tabular}
}
}
\end{table}

\section{Conclusion}
\label{sec:majhead}

In this paper, we propose a fully time-domain improved feature extraction network for neuro-oriented target speaker extraction. Notably, we introduce SpeechBiMamba, which specializes in long sequence modeling of speech signals, and EEGKAN which is used for EEG coding to locate the target speaker through the subject's attention information. Experimental outcomes underscore the significant contribution of EEG features, extracted via the EEGKAN layer, in facilitating attentive speaker extraction. Furthermore, the proposed IFENet methodology demonstrates remarkable efficacy in extracting target speech without relying on any prior knowledge about the target speaker, thereby showcasing its robustness and applicability. In the future, we plan to explore more modal information for speaker extraction, including but not limited to EEG signals and observable lip movements.

\vspace{12pt}
\clearpage

\end{document}